\documentclass[aps,pra,groupedaddress,twocolumn]{revtex4-2}

\usepackage{graphicx}
\usepackage{amssymb}
\usepackage{amsmath}
\usepackage{bm}
\usepackage{color}
\usepackage{physics}

\usepackage[colorlinks,urlcolor=blue,citecolor=blue,linkcolor=blue]{hyperref}
\usepackage{comment}
\usepackage{cleveref}
\usepackage{mathtools}
\usepackage{mathrsfs}

\newcommand{\eb}{\varepsilon_B}
\newcommand{\nn}{\nonumber}
\renewcommand{\k}{\mathbf{k}}
\newcommand{\ek}{\epsilon_\k}
\newcommand{\0}{\mathbf{0}}
\newcommand{\p}{\mathbf{p}}
\newcommand{\q}{\mathbf{q}}
\newcommand{\Q}{\mathbf{Q}}

\newcommand{\PP}{\operatorname{PP}}
\newcommand{\PX}{\operatorname{PX}}

\newcommand{\kP}{\mathbf{k}^\prime}

\newcommand{\diff}{\mathop{}\!\mathrm{d}}

\begin{document}

\title{Microscopic calculation of polariton scattering in 
semiconductor microcavities}

\author{Guangyao Li}
\affiliation{School of Physics and Astronomy, Monash University, Victoria 3800, Australia}
\affiliation{ARC Centre of Excellence in Future Low-Energy Electronics Technologies, Monash University, Victoria 3800, Australia}

\author{Meera M. Parish}
\affiliation{School of Physics and Astronomy, Monash University, Victoria 3800, Australia}
\affiliation{ARC Centre of Excellence in Future Low-Energy Electronics Technologies, Monash University, Victoria 3800, Australia}

\author{Jesper Levinsen}
\affiliation{School of Physics and Astronomy, Monash University, Victoria 3800, Australia}
\affiliation{ARC Centre of Excellence in Future Low-Energy Electronics Technologies, Monash University, Victoria 3800, Australia}

\date{\today}

\begin{abstract}
    Recent experiments in exciton-polariton systems have provided high-precision measurements of the value of the polariton-polariton interaction constant,  which is a key parameter that governs the nonlinear dynamics of polariton condensates and potentially enables quantum correlated polaritons. Yet, until now, this parameter has only been addressed  theoretically using perturbative treatments or approximations that do not include the composite nature of the excitons. Here, we use a recently developed microscopic description of polaritons involving electrons, holes, and photons, where the interactions between charged particles are assumed to be highly screened. Within this model, we perform an exact four-body calculation of the spin-polarized  polariton-polariton and polariton-exciton interaction constants. 
    In the limit of weak light-matter coupling relevant to an atomically thin semiconductor in a microcavity, we obtain excellent agreement with a recently proposed universal form of low-energy polariton-polariton scattering [O. Bleu \textit{et al.}, Phys. Rev. Res. \textbf{2}, 043185 (2020)]. At stronger light-matter coupling, of relevance to multilayer microcavities, we observe that the interaction constant increases towards that predicted by the Born approximation. We show that in all regimes of interest the interaction constant can be accurately obtained from the exciton-exciton scattering phase shift at negative collision energy, and we argue that this has important implications for interactions in other systems featuring strong light-matter coupling.
\end{abstract}

\maketitle

\section{Introduction}

Exciton-polaritons are hybrid light-matter quasiparticles resulting from the strong coupling between photons trapped inside a microcavity and excitons in a two-dimensional (2D) semiconductor 
\cite{DengRevMod10,CiutiRevMod13}. 
Such microcavity polaritons provide a powerful platform for the investigation of Bose-Einstein condensation~\cite{BEC06,BaliliScience07}, superfluidity~\cite{Amo2009,Sanvitto2010,LerarioNatPhys2017}, polariton lasers \cite{Deng2003,ChristopoulosPRL2007,SchneiderNature2013}, and topological phase transitions~\cite{Solnyshkov2021}. 
Many of these applications rely on 
the interactions within a polariton condensate, or the interaction between condensed polaritons and an excitonic reservoir. In particular, the nonlinear dynamics of a polariton condensate has been 
successfully modelled by an open-dissipative Gross–Pitaevskii equation \cite{WouterPRL07} with nonlinear interaction coefficients given by the polariton-polariton 
and polariton-exciton 
interaction constants $g_{\PP}$ and $g_{\PX}$, respectively. 
Both of these are therefore of vital importance for an accurate modelling of polaritonic systems. While experimental measurements have historically differed from each other by several orders of magnitude \cite{EliPRB2019,snoke2021REV}, recently there have been several high-precision measurements which have narrowed down the possible range of interaction constants~\cite{EliPRB2019,Munoz2019,Delteil2019,MaciejNatComm2020,
Stepanov2021}.

Microscopically, the polariton nonlinearity originates from their excitonic component. This has motivated the widely used expressions for the polariton-polariton and polariton-exciton interaction constants~\cite{TassonePRB99}:
\begin{align}
g_{\rm PP}^{\rm Born} =|X_{\rm LP}|^4g_{\rm XX}^{\rm Born},\qquad g_{\rm PX}^{\rm Born}=|X_{\rm LP}|^2g_{\rm XX}^{\rm Born}.
\label{eq:Born}
\end{align}
Here, $|X_{\rm LP}|^2$ is the excitonic fraction of the lower polariton, and $g_{\rm XX}^{\rm Born}$ is the exciton-exciton interaction constant calculated within the Born approximation, assuming spin polarized polaritons. While there have been several proposed corrections to these expressions, see, e.g., Refs.~\cite{TassonePRB99,RochatPRB00,Combescot_2007,GlazovPRB2009,Brichkin2011}, until recently these have all been perturbative in nature. This lack of progress towards determining these key parameters of polariton physics is primarily due to the complexity of including multiple electronic exchange processes, and the challenge of taking into account the modification of the internal structure of the exciton due to the coupling to light.

Recently, we have demonstrated that the polariton interaction constant can be calculated exactly in the limit of weak light-matter Rabi coupling relative to the $1s$ exciton binding energy~\cite{Bleu2020}. This result is, for instance, of direct applicability to the emerging class of atomically thin transition metal dichalcogenide (TMD) semiconductors, which feature tightly bound excitons. Generalizing our result~\cite{Bleu2020} to the polariton-exciton interaction constant, we have the following universal expressions for a single semiconductor layer
\begin{align}\label{eq:Olivier_expression}
    g_{\PP}=\frac{4\pi \hbar^2\abs{X_{\rm LP}}^4}{m_X\ln\left(\frac{\varepsilon_X}{2\abs{E_{\rm LP}+\eb}}\right)},
    \quad g_{\rm PX}=\frac{4\pi \hbar^2\abs{X_{\rm LP}}^2}{m_X\ln\left(\frac{\varepsilon_X}{\abs{E_{\rm LP}+\eb}}\right)},
\end{align}
where $E_{\rm LP}$ is the lower polariton energy measured from the electron-hole band gap, and $m_X$ and $\eb$ are the 1$s$ exciton mass and binding energy, respectively.
Importantly, just like the Born approximation in Eq.~\eqref{eq:Born}, these expressions only depend on a single parameter, $\varepsilon_X$, that characterizes the excitonic interactions.
Furthermore, since $\varepsilon_X$ appears under a logarithm, the resulting interaction constant is quite insensitive to its precise value and thus $\varepsilon_X$ may simply be replaced by the exciton binding energy to a good approximation.
Low-energy expressions such as Eq.~\eqref{eq:Olivier_expression} have already provided a possible explanation~\cite{Bleu2021} for the size of the polariton antibunching observed in recent experiments~\cite{Delteil2019}. 
It has also been shown that $g_{\rm PP}$ in Eq.~\eqref{eq:Olivier_expression} emerges in a Gaussian pair fluctuation treatment of a polariton condensate~\cite{HUPRA2020}.

Equation~\eqref{eq:Olivier_expression} was derived within a model that assumes structureless (tightly bound) excitons~\cite{Bleu2020}. In order to go beyond this to systems that feature larger light-matter coupling relative to the exciton binding energy --- such as GaAs quantum wells or multilayer TMD systems --- where the internal structure of the excitons becomes important, one needs to introduce the constituent electrons and holes explicitly into the theory. The description of a polariton as a superposition of a photon and an electron-hole pair has only recently been developed, first in a model featuring Coulomb electronic interactions~\cite{LevinsenPRR19} and later in the case of strongly screened interactions~\cite{Hu2020QuantumFI,Li2021PRL,Li2021PRB}. These models can provide the basis of few-body scattering calculations in exciton-polariton systems. 
However, while there has been recent progress in exactly solving the three-body problem with Coulomb interactions~\cite{CombescotPRX2017,Fey2020}, 
such calculations are highly challenging and have not yet been carried out for the four-body problem (two electrons and two holes) in the presence of coupling to light. 

In this work, we use a diagrammatic approach to perform a full four-body calculation of the polariton-polariton and polariton-exciton scattering in a 2D semiconductor microcavity. The calculation is based on the electron-hole-photon polariton model with strongly screened Coulomb interactions, which we have already successfully used to derive universal features of polariton-electron scattering~\cite{Li2021PRL,Li2021PRB}. 
Within the accuracy of the model, the diagrammatic approach sums all contributions to the interaction constants, and it is thus exact. Specifically, it includes all possible direct and exchange processes present in the model as well as the saturation of the exciton oscillator strength due to strong light-matter coupling. In the case of small light-matter coupling, our calculations perfectly match Eq.~\eqref{eq:Olivier_expression}, providing further evidence for this universal low-energy description of polariton interactions. Whereas this simplified expression loses its validity when the Rabi coupling is large, our full four-body calculation does not suffer from this drawback. 
In particular, we find that the interaction constants approach the predicted results from the Born approximation \eqref{eq:Born} in the regime of a very strong light-matter coupling that exceeds the exciton binding energy.
Thus, our approach allows us to explore the full range of light-matter coupling strengths, from the universal behavior at weak Rabi coupling towards the perturbative regime at very strong coupling. Finally, we find that a simple generalization of Eq.~\eqref{eq:Olivier_expression} based on the exciton-exciton scattering phase shift at negative collision energy can accurately reproduce our calculated interaction constants, a result which has general implications for strongly coupled light-matter systems.

This paper is organized as as follows. In Sec.~\ref{sec:Hamiltonian}, we present the model Hamiltonian and review its properties including the theoretical description of the resulting polariton states. In Sec.~\ref{sec:PP}, we discuss in detail the full set of four-body diagrams for the polariton-polariton scattering process, their numerical solution, and the results for the interaction constant. In Sec.~\ref{sec:PX}, we extend our diagrammatic approach to the case of polariton-exciton interactions, and in Sec.~\ref{sec:multi} we discuss the extension of our theory to multilayer systems, and the comparison with experimental data.
In Sec.~\ref{sec:conclusion} we conclude. Technical details are given in the Appendices.

\section{Hamiltonian}\label{sec:Hamiltonian}

\subsection{Model}\label{subsec:polariton_states}

We consider a 2D semiconductor embedded in a  
microcavity. The semiconductor can either be an atomically thin TMD~\cite{TMDREV2018} or a conventional quantum well such as GaAs \cite{KavokinMicrocavity}. Our aim 
is to calculate the scattering of identical polaritons, and therefore we will throughout consider a spin-polarized system. 
For simplicity, in the following we consider only bright excitons that directly couple to light, and we 
specialize to a single active layer; however our results also apply with minor modifications to multilayer systems that feature both bright and dark excitons, as discussed in Sec.~\ref{sec:multi}. 
Approximating the interaction between charged particles by a contact interaction, we therefore use the Hamiltonian
\begin{align}\label{eq:Hamiltonian}
\hat{H}=&\sum_\k \left(\epsilon^e_\k e^\dagger_\k e_\k+\epsilon^h_\k h^\dagger_\k h_\k \right) - V_0 \sum_{\k \k'\q} e^\dagger_{\k}h^\dagger_{\q-\k}h_{\q-\kP}e_{\kP} \nn \\
&+\sum_\k(\omega+\epsilon^c_{\k})c^\dagger_{\k}c_{\k} 
+g\sum_{\k\,\q}\left( e^\dagger_{\k} h^\dagger_{\q-\k} c_{\q} +
\rm{H.c.} 
\right),
\end{align}
where we work in units where $\hbar$ and the system area are set to unity. Here, the first line describes the semiconductor, while the second line describes the coupling to a single photonic mode inside the microcavity.
$e^\dag_\k$, $h^\dag_\k$, and $c^\dag_\k$ are creation operators with planar momentum $\k$ for electron, hole, and photon, respectively. They have the corresponding dispersions $\epsilon^{e,h,c}_\k=\k^2/2m_{e,h,c}$ with effective masses $m_{e,h,c}$, where we have separated out the zero-momentum cavity photon energy in the absence of light-matter coupling, $\omega$. 
The contact electron-hole interaction of strength $-V_0$ serves as an approximation of the highly screened Coulomb interaction, where the minus sign implies an attractive interaction. 
On the other hand, owing to Pauli exclusion, electrons (and holes) do not interact among themselves. The last term describes the light-matter coupling of strength $g$, where we have applied the rotating  wave  approximation.
Note that all energies are defined with respect to the semiconductor band gap. 

The coupling of light and matter in the semiconductor microcavity leads to the formation of exciton polaritons. These new quasiparticles are often described using a model of two coupled oscillators consisting of the cavity photon mode and the exciton~\cite{CiutiRevMod13}. Taking the exciton-photon coupling to have strength $\Omega$ and the exciton energy at momentum $\k$ to be $-\eb+\ek^X$, with $\ek^X=\k^2/2m_X$ the exciton dispersion and $m_X=m_e+m_h$ the exciton mass, we have the coupled-oscillator Hamiltonian
\begin{align}
    \hat{H}^{\rm osc} = 
    \begin{pmatrix}
    -\eb+\ek^X & \Omega \\
    \Omega & \omega + \epsilon^c_{\k}
    \end{pmatrix} .
\end{align}
Solving for the eigenstates, this then gives the upper $(+)$ and lower $(-)$ polariton dispersions
\begin{align}  \label{eq:Eosc}
E^{\rm osc}_\pm(\k)  = & -\eb + \frac12 \left(\delta^{\rm osc}+\epsilon^c_\k+\epsilon^X_\k \right) \nn \\
 & \pm \frac12 \sqrt{(\delta^{\rm osc}+\epsilon^c_\k-\epsilon^X_\k)^2+4\Omega^2} .
\end{align}
Here we have defined the photon-exciton detuning 
at normal incidence, $\delta^{\rm osc}\equiv\omega+\eb$. 
Within the coupled oscillator model, the polaritons have the corresponding photon fraction
\begin{align}
    |C^{\rm osc}_{\pm}(\k)|^2&=\frac{1}{2}\left(1\mp\frac{\epsilon^X_\k-\delta^{\rm osc}-\epsilon^c_\k}{\sqrt{\left(\epsilon^X_\k-\delta^{\rm osc}-\epsilon^c_\k\right)^2+4\Omega^2}}\right),
    \label{eq:Cosc}
\end{align}
and exciton fraction $|X^{\rm osc}_{\pm}(\k)|^2=1-|C^{\rm osc}_{\pm}(\k)|^2$. The coefficients $X$, $C$ are also known as Hopfield coefficients.

The model employed in this work is obviously more complicated, as it explicitly includes electrons and holes rather than a structureless exciton. In particular, the use of contact light-matter and electron-hole interaction potentials introduces ultraviolet divergences that must be cured by the process of renormalization. This was first carried out in the case of Coulomb electron-hole interactions in Ref.~\cite{LevinsenPRR19}, and for contact electron-hole interactions in Ref.~\cite{Hu2020QuantumFI}. In the following we use the scheme of Refs.~\cite{Li2021PRB,Li2021PRL} which has the advantage that it is fully analytic. 

The Hamiltonian~\eqref{eq:Hamiltonian} features an electron-hole bound state with the appropriate dispersion, which we identify as the $1s$ exciton state. In terms of the parameters of the model, the exciton binding energy satisfies (see, e.g., Ref.~\cite{LevinsenBook15})
\begin{align}
    \frac1{V_0}=\sum_\k^\Lambda \frac1{\eb+\ek^e+\ek^h},
\end{align}
where $\Lambda$ is an ultraviolet cutoff. Applying the same cutoff to the light-matter interactions, we can relate the physically meaningful Rabi coupling to the bare model parameters $V_0$ and $g$ via \cite{Li2021PRL,Li2021PRB}:
\begin{align}\label{eq:Rabi_renormal}
    \Omega=\frac{g}{V_0}\sqrt{\frac{2\pi\eb}{m_r}},
\end{align}
where $m_r^{-1}=m_e^{-1}+m_h^{-1}$ is the electron-hole reduced mass and we take the limit $\Lambda\to\infty$. The cavity photon-exciton detuning within this model is
\begin{align}
    \delta = \omega+\eb-\frac{\Omega^2}{2\eb}.
\end{align}

We then obtain the polariton spectrum by solving the implicit equation for the energy $E$~\cite{Li2021PRB}
\begin{align}
    \left(\delta-\eb+\frac{\Omega^2}{2\eb}+\epsilon_\k^c-E\right)\ln\left[\frac{\epsilon_\k^X-E}{\eb} \right]=\frac{\Omega^2}{\eb},
    \label{eq:Efull}
\end{align}
which yields two states below the electron-hole continuum that starts at the bandgap energy~\footnote{The exception to this statement is cases where we have a very large Rabi coupling and/or a large positive detuning, when instead the upper polariton exists as a resonance in the continuum}.
Defining the energy of these two states to be $E_\pm(\k)$, the associated photon fractions are
\begin{align}
    |C_\pm(\k)|^2=\frac1{1+
    \frac{\left(E_\pm(\k)-\delta+\eb-\frac{\Omega^2}{2\eb}-\ek^c\right)^2}{\Omega^2}\frac{\eb}{|E_\pm(\k)-\ek^X|}},
    \label{eq:Cfull}
\end{align}
and $|X_\pm(\k)|^2=1-|C_\pm(\k)|^2$.
Reference~\cite{Li2021PRB} demonstrated that Eqs.~\eqref{eq:Efull} and \eqref{eq:Cfull} very accurately match the results of the simple coupled-oscillator model in Eqs.~\eqref{eq:Eosc} and \eqref{eq:Cosc} for Rabi couplings up to $\Omega\sim\eb$.

Before moving on to utilize the above results, we discuss the relationship between our Hamiltonian in Eq.~\eqref{eq:Hamiltonian} and realistic models of materials.
In practice, electrons and holes interact with each other via an unscreened or partially-screened Coulomb potential --- for instance the Keldysh
potential~\cite{Rytova1967, Keldysh1979,CudazzoPRB2011} --- which is long-range. By contrast, the contact interaction used here might appear as an oversimplification. Nevertheless, recent theoretical calculations on the exact exciton-electron scattering in atomically thin semiconductors \cite{Fey2020} involving the Coulomb/Keldysh potentials have obtained a phase shift that has the same universal behavior at low collision energies as that obtained from the Hamiltonian with contact interactions \cite{VudtiwatEPL13,Li2021PRB}.
Physically, this is due to the fact that exciton-electron interactions are of the charge-induced dipole type and are therefore short range \cite{EfimkinPRB2018}. This reasoning can be extended to the case of polariton-polariton interactions where the underlying exciton-exciton interaction is that of two induced dipoles, which is even shorter ranged than exciton-electron interactions. 
Indeed, as we will see below, the full four-body calculation matches the universal low-energy behaviour of two-particle scattering presented in Eq.~\eqref{eq:Olivier_expression}, in agreement with previous microscopic calculations of exciton-exciton scattering~\cite{Schindler2008} but in contrast to the Born approximation \cite{TassonePRB99}.

\subsection{Foundations of diagrammatic approach}

In the diagrammatic approach to few-body problems involving polaritons~\cite{Li2021PRL,Li2021PRB}, we first need to define the relevant propagators. The single-particle electron and hole propagators at momentum $\k$ and energy $E$ take the form
\begin{align}
    G_{e,h}(\k,E)=\frac1{E-\ek^{e,h}+i0},
\end{align}
where in the denominator the infinitesimal $+i0$ shifts the energy pole slightly into the lower half of the complex $E$ plane indicating the retarded Green's function \cite{FetterBook}.

By considering the sum of all possible repeated interactions between an electron and a hole in the absence of light-matter coupling, we arrive at the electron-hole $T$ matrix at total momentum $\k$ and energy $E$~\cite{LevinsenBook15} 
\begin{align}\label{eq:D0}
    D_0(\k,E)=\frac{2\pi/m_r}{-\ln\left[\frac{\ek^X-E-i0}{\eb}\right]},
\end{align}
the pole of which precisely corresponds to the exciton energy. We will refer to this as the (unnormalized) exciton propagator.
Including light-matter coupling, Ref.~\cite{Li2021PRB} found that the electron-hole $T$ matrix becomes
\begin{align}\label{eq:Dp}
    D(\k,E)=\frac{2\pi/m_r}{-\ln\left[\frac{\ek^X-E-i0}{\eb}\right]-\frac{\Omega^2}{\eb}\frac1{E-\delta+\eb-\frac{\Omega^2}{2\eb}-\ek^c+i0}}.
\end{align}
To distinguish Eq.~\eqref{eq:Dp} from the electron-hole $T$ matrix in the absence of light-matter coupling, we will refer to this as the polariton propagator.

The denominator of the polariton propagator in Eq.~\eqref{eq:Dp} gives rise to two energy poles corresponding precisely to the dispersions of upper and lower polaritons in Eq.~\eqref{eq:Efull}. The residue of the propagator at the poles gives the normalization factors~\cite{Li2021PRB}
\begin{align}\label{eq:Dnorm}
    Z_\pm(\k)|X_\pm(\k)|^2,
\end{align}
where we define
\begin{align}
    Z_\pm(\k)=\frac{2\pi|E_\pm(\k)-\ek^X|}{m_r}.
\end{align}
We see that the normalization in Eq.~\eqref{eq:Dnorm} is directly related to the polariton exciton fraction~\cite{Li2021PRB}. 

In the remainder of this paper, we will restrict our attention to the lower polariton branch, which is typically probed in experiments investigating polariton interactions. For convenience, we will therefore denote the polariton energy, exciton Hopfield coefficient, and normalization at zero in-plane momentum by $E_{\rm LP}=E_-(0)$, $X_{\rm LP}=X_-(0)$, and $Z_{\rm LP}=Z_-(0)$.

\section{Polariton-Polariton scattering}
\label{sec:PP}

We now proceed to solve the polariton-polariton (P-P) scattering problem exactly within the model introduced in Sec.~\ref{sec:Hamiltonian}. We take advantage of the diagrammatic techniques that were first developed for three-body problems in nuclear physics~\cite{Skorniakov1957}, and later extended to the four-body problem in the context of cold atomic gases~\cite{Brodsky2005, LevinsenPRA06, BrodskyPRA06,Levinsen2011}. Indeed, the model Hamiltonian in cold atoms is precisely the same as the first line of Eq.~\eqref{eq:Hamiltonian}, i.e., the electron-hole problem with strongly screened interactions and no coupling to light. For the case of equal-mass atoms (carriers), the 2D low-energy dimer-dimer (exciton-exciton) scattering has previously been calculated in Ref.~\cite{Petrov2003} using an alternative wave function based approach. 

In the following, we will first outline the diagrammatic calculation and then we will present the results both at weak and at strong light-matter coupling relative to the exciton binding energy. Finally, we will discuss how an extended version of the universal low-energy expression in Eq.~\eqref{eq:Olivier_expression} allows us to interpret our results across the whole range of light-matter coupling strengths.

\begin{figure}[t!]
	\centering
	\includegraphics[width=\linewidth]{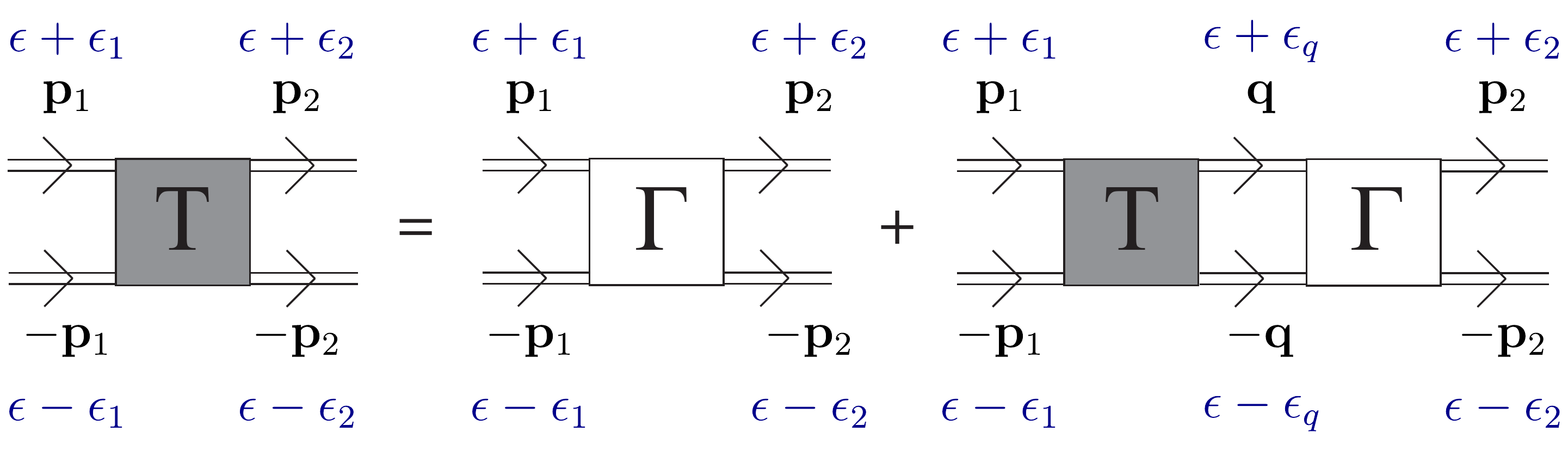}
	\caption{Diagrammatic representation of Eq.~\eqref{eq:T-matrix_full} satisfied by the polariton-polariton scattering $T$ matrix (shaded square).  
	$\Gamma$ (white square) is the sum of all two-polariton-irreducible diagrams. The polariton propagator in Eq.~\eqref{eq:Dp} is represented by the double-line arrow. The corresponding scattering momentum and energy are listed adjacent to each polariton propagator with $E$ the total energy, $\epsilon=E/2$ the energy per particle, and $\epsilon_{1,2}$ the energy difference for the incoming and outgoing scattering pairs, respectively. Note that external propagators are not included in $T$ or $\Gamma$. }
	\label{fig:PP_T-matrix}
\end{figure}

\subsection{Four-body formalism}\label{subsec:four_body_diagrams}

The P-P scattering process can be readily 
calculated by extending the diagrammatic approach developed for the scattering of diatomic Feshbach molecules in an ultracold atomic gas~\cite{Brodsky2005, LevinsenPRA06, BrodskyPRA06,Levinsen2011} to include the polariton propagator in Eq.~\eqref{eq:Dp}. The fundamental difference is the following.
In standard scattering theory, a scattering process is calculated in the center-of-mass frame of the interacting pair, where a partial wave expansion can be performed in which the partial waves decouple~\cite{TaylorBook}.
In semiconductor microcavities, the light-matter coupling results in a nonparabolic polariton dispersion. This is a signature of a broken Galilean invariance, that prevents the use of a Galiliean transformation between the laboratory frame and the center-of-mass frame. 
The issue of finding a reliable reference frame transformation for a superposition of two quantum systems of different mass is nontrivial \cite{GiacominiNatComm19}, and likewise performing the calculation in the laboratory frame where all partial waves are intrinsically coupled is not an easy task. 
To circumvent this difficulty, we restrict our calculations to the special situation where the laboratory frame coincides with the center-of-mass frame, i.e., where incoming and outgoing polaritons have equal and opposite momenta.

The strength of the corresponding interaction is quantified by a scattering $T$ matrix which has contributions from an infinite number of diagrams. However, fortunately these can be exactly resummed by solving a set of integral equations~\cite{Brodsky2005, LevinsenPRA06, BrodskyPRA06,Levinsen2011}. In this work, we will use the formulation of Ref.~\cite{Levinsen2011}. Figure~\ref{fig:PP_T-matrix} illustrates diagrammatically the integral equation satisfied by the P-P scattering $T$ matrix, where the external incoming and outgoing polaritons determine the scattering parameters such as energy and momentum, but otherwise do not form part of the $T$ matrix. Note the similarity between the diagrams in Fig.~\ref{fig:PP_T-matrix} and the ladder approximation to the Bethe-Salpeter equation~\cite{FetterBook}.
To be explicit, we consider the scattering of two identical polaritons with opposite initial and final momenta $\pm\p_1$ and $\pm\p_2$, respectively, and we denote the $T$ matrix by $T(E;\p_1,\epsilon_1;\p_2, \epsilon_2)$, where $E$ is the total collision energy, and the energy of incoming (outgoing) polaritons is shifted by $\pm\epsilon_1$ ($\pm\epsilon_2$) relative to the energy per particle, $\epsilon=E/2$. 
The low-energy P-P interaction constant is then obtained by taking the total energy $E=2E_{\rm LP}$, while $\p_i$ and $\epsilon_i$ are set to 0:
\begin{align}
    g_{\rm PP}=T(2E_{\rm LP};\0,0;\0,0).
    \label{eq:gPP}
\end{align}
This is the interaction constant that would appear, for instance, in a Gross-Pitaevskii equation treatment of a dilute polariton condensate~\cite{Pitaevskiibook,CiutiRevMod13}. It is therefore a key parameter in polariton physics.

The block $\Gamma$ appearing on the right hand side of the equation in Fig.~\ref{fig:PP_T-matrix}  represents the sum of all possible two-polariton-irreducible processes~\footnote{Diagrams that cannot be cut in half at any intermediate step by cutting two polariton propagators only.} that begin and end with two polaritons. The diagrammatic representation has a one-to-one correspondence with the following integral equation for the unnormalized $T$ matrix:
\begin{widetext}
\begin{align}
	t(E;\p_1,\epsilon_1;\p_2, 
	\epsilon_2)=&\Gamma(\p_1,\epsilon_1;\p_2, \epsilon_2)\nonumber\\
													&+
													i\int \frac{d\epsilon_q}{2\pi}\sum_\q t(E; \p_1,\epsilon_1;\q, \epsilon_q)D (\q,\epsilon+\epsilon_q)D(\q,\epsilon-\epsilon_q)\Gamma(\q,\epsilon_q;\p_2,\epsilon_2),\label{eq:T-matrix_full}
\end{align}
where we suppress the dependence of $\Gamma$ on $E$.
The full $T$ matrix is then given by
\begin{align}\label{eq:Tfromt}
    T(E;\p_1,\epsilon_1;\p_2, 
	\epsilon_2)=Z_-(\p_1)Z_-(\p_2)|X_-(\p_1)|^2|X_-(\p_2)|^2t(E;\p_1,\epsilon_1;\p_2, 
	\epsilon_2)
\end{align}
where the prefactor is the normalization factor in Eq.~\eqref{eq:Dnorm} that accounts for the residue of the four external polariton propagators \cite{Li2021PRB}. Note that Eq.~\eqref{eq:T-matrix_full} depends parametrically on $\p_1$ and $\epsilon_1$, and therefore it can be solved separately for each value of these.

\begin{figure}[t!]
	\centering
	\includegraphics[width=.5\linewidth]{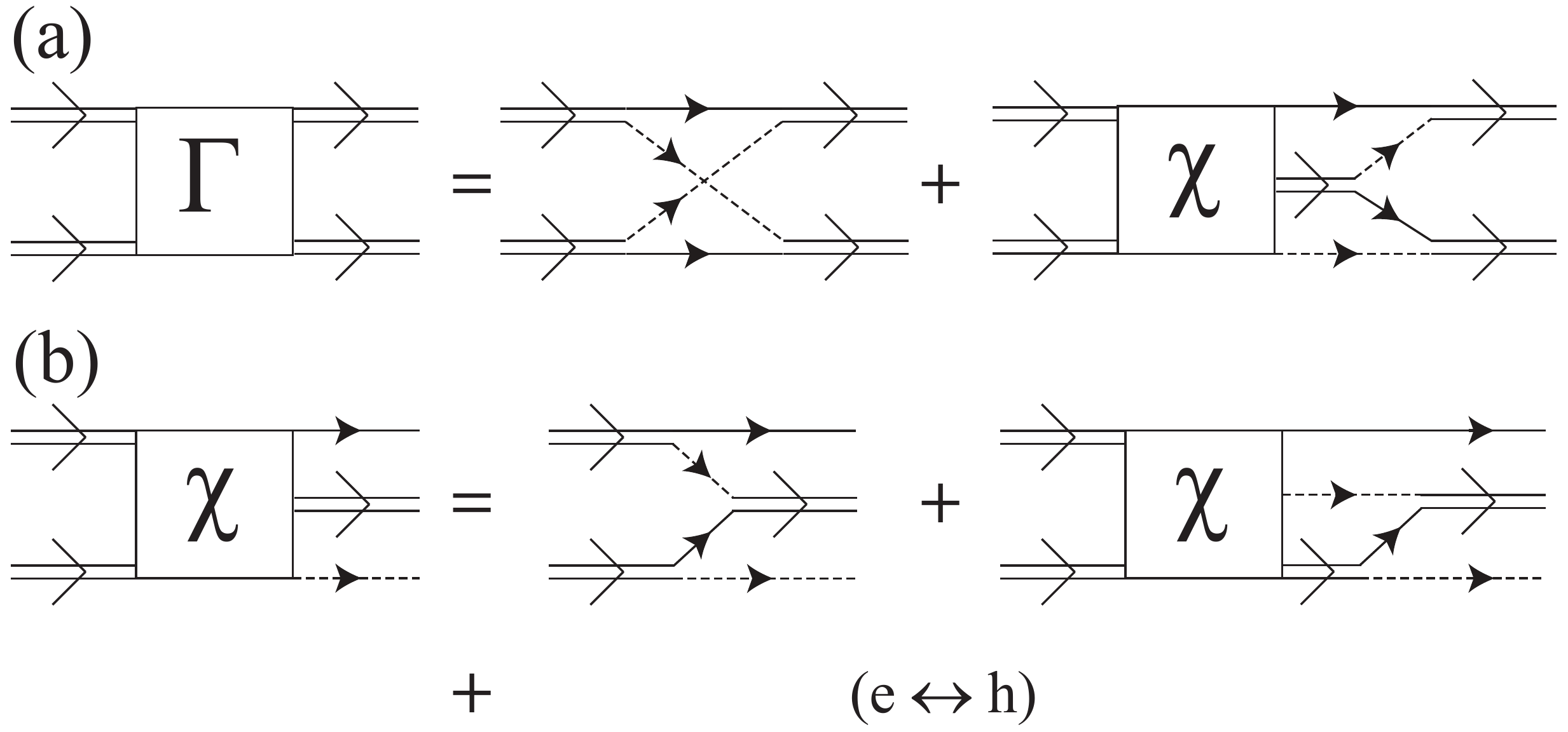}
	\caption{Diagrammatic representation of the $\Gamma$ and $\chi$ functions. (a) $\Gamma$ consists of a simple fermion exchange process and the infinitely repeated polariton decomposition process represented by $\chi$. (b) The integral equation satisfied by $\chi$. It represents a polariton decomposing and recombining process such that one of the outgoing polaritons is split into a free electron and hole. The last line represents the same set of diagrams with electron and hole propagators being swapped. The solid- and dashed-arrow lines represent the propagators of a free electron and a free hole, respectively.}
	\label{fig:Gamma_and_chi}
\end{figure}

Next, we proceed with the construction of $\Gamma$ that appears in the kernel of our integral equation. This is conveniently written as the sum of two 
diagrams: a simple exchange of the two fermions, denoted as $\Gamma^{(0)}$; and another describing the processes where we have formation of an intermediate polariton. 
The construction of $\Gamma$ is illustrated in Fig.~\ref{fig:Gamma_and_chi}(a). The $\chi$ block represents the sum of all two-polariton-irreducible processes that begin with two polaritons and end with one polariton and two free fermions. From Fig.~\ref{fig:Gamma_and_chi}(a), the expression for $\Gamma$ can be written as:
\begin{align}
    \Gamma(\p_1,\epsilon_1;\p_2,\epsilon_2)=\Gamma^{(0)}(\p_1,\epsilon_1;\p_2,\epsilon_2)-\frac{1}{2}\sum_{\Q_1\Q_2}
    &\left[G_h(\p_2-\Q_1,\epsilon+\epsilon_2-\epsilon^e_{Q_1}  )G_e(\p_2+\Q_2,\epsilon-\epsilon_2-\epsilon^h_{Q_2})\right.\nn\\
    &\left.+G_h(\p_2+\Q_1,\epsilon-\epsilon_2-\epsilon^e_{Q_1})G_e(\p_2-\Q_2,\epsilon+\epsilon_2-\epsilon^h_{Q_2})           \right]\nn\\
    &\times D(\Q_1+\Q_2,2\epsilon-\epsilon^e_{Q_1}-\epsilon^h_{Q_2})\chi(\p_1,\epsilon_1;\Q_1,\Q_2).\label{eq:Gamma_full}
\end{align}
Here, $\Q_1$ and $\Q_2$ are the internal momenta within the fermion exchanged loops emitting on the right of the $\chi$ function in Fig.~\ref{fig:Gamma_and_chi}(a) and we have already carried out the corresponding energy integration by Cauchy’s residue theorem \cite{Levinsen2011}. Again, we suppress the dependence of $\chi$ on $E$, and $Q_{1,2}=\abs{\Q_{1,2}}$ is the magnitude of the vector.

\begin{figure}[h]
	\centering
	\includegraphics[width=.5\linewidth]{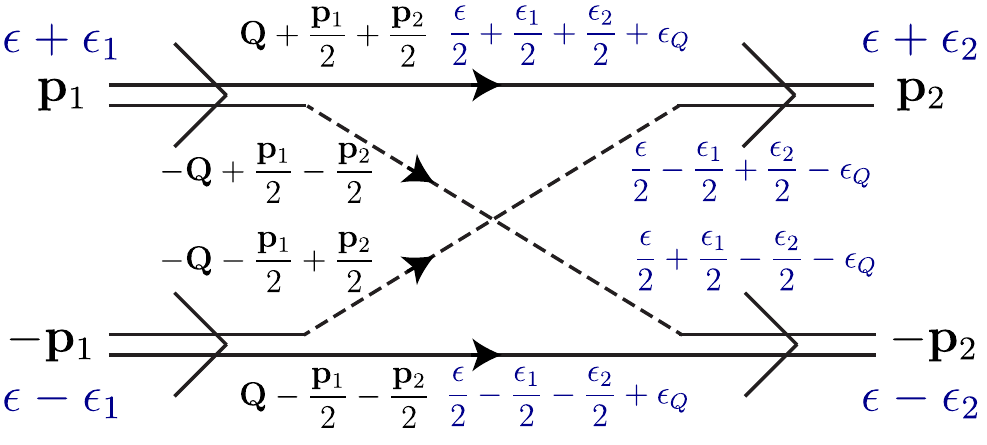}
	\caption{Diagrammatic representation of the $\Gamma^{(0)}$ function with symmetric energy and momentum exchanged.}
	\label{fig:Gamma(0)}
\end{figure}

Figure~\ref{fig:Gamma(0)} shows in detail the energy and momentum exchanged within $\Gamma^{(0)}$. 
In Fig.~\ref{fig:Gamma(0)}, we adopt a symmetric labelling of each of the exchanged energy and momentum for the convenience of later integration of $\epsilon_Q$ and $\Q$. Written out explicitly, $\Gamma^{(0)}$ reads:
\begin{align}
	\Gamma^{(0)}(\p_1,\epsilon_1; \p_2,\epsilon_2)=-i\int \frac{\diff \epsilon_Q }{2\pi} \sum_{\Q}& \left[ G_e(\Q+\frac{\p_1}{2}+\frac{\p_2}{2},\frac{\epsilon}{2}+\frac{\epsilon_1}{2}+\frac{\epsilon_2}{2}+\epsilon_Q)
	G_e(\Q-\frac{\p_1}{2}-\frac{\p_2}{2},\frac{\epsilon}{2}-\frac{\epsilon_1}{2}-\frac{\epsilon_2}{2}+\epsilon_Q)\right.\nonumber\\
	&\left.\times G_h(-\Q+\frac{\p_1}{2}-\frac{\p_2}{2},\frac{\epsilon}{2}+\frac{\epsilon_1}{2}-\frac{\epsilon_2}{2}-\epsilon_Q)
	G_h(-\Q-\frac{\p_1}{2}+\frac{\p_2}{2},\frac{\epsilon}{2}-\frac{\epsilon_1}{2}+\frac{\epsilon_2}{2}-\epsilon_Q)\right],\label{eq:Gamma(0)_ori}
\end{align}
where the minus sign comes from the exchange of identical fermions. 
The energy integration of $\epsilon_Q$ over the real axis can be performed by either enclosing the upper-half or lower-half complex $\epsilon_Q$ plane and summing over the residues of $2$ simple poles. We provide an explicit expression in Appendix~\ref{app:appendix}.

Now we proceed to investigate the $\chi$ function. 
By definition, this contains all possible two-polariton-irreducible processes that end up with one polariton and two free fermions. 
The sum of those infinite processes again allows us to write down an integral equation for $\chi$ that automatically includes all orders. Figure~\ref{fig:Gamma_and_chi}(b) shows the diagrammatic representation of the integral equation, which reads:
\begin{align}
    &\chi(\p_1,\epsilon_1;\Q_1,\Q_2)=-\left[G_h(\p_1-\Q_1,\epsilon+\epsilon_1-\epsilon^e_{Q_1}) G_e(\p_1+\Q_2,\epsilon-\epsilon_1-\epsilon^h_{Q_2})+ [(\p_1,\epsilon_1)\leftrightarrow(-\p_1,-\epsilon_1)]
    \right] \nn \\
    &-\sum_{\Q_3}
    \left[G_e(\Q_1+\Q_2+\Q_3,2\epsilon-\epsilon^e_{Q_1}-\epsilon^h_{Q_2}-\epsilon^h_{Q_3})D(\Q_1+\Q_3,2\epsilon-\epsilon^e_{Q_1}-\epsilon^h_{Q_3})\chi(\p_1,\epsilon_1;\Q_1, \Q_3)\right.\nn\\
    &\qquad\quad\left.+G_h(\Q_1+\Q_2+\Q_3,2\epsilon-\epsilon^e_{Q_1}-\epsilon^h_{Q_2}-\epsilon^e_{Q_3})D(\Q_2+\Q_3,2\epsilon-\epsilon^h_{Q_2}-\epsilon^e_{Q_3})\chi(\p_1,\epsilon_1;\Q_3, \Q_2)   \right].\label{eq:chi_full}
\end{align}
Like Eq.~\eqref{eq:T-matrix_full}, this equation depends parametrically on $\p_1$ and $\epsilon_1$.
\end{widetext}

In the following, we will also compare our results with the Born approximation. This corresponds to replacing the full sum of diagrams in the $T$ matrix by the fermion exchange diagram $\Gamma^{(0)}$ in Fig.~\ref{fig:Gamma_and_chi}(a). 
This allows us to calculate the corresponding approximation of the interaction constant
\begin{align}
    \label{eq:gPPBorn}
    g_{\rm PP}^{\rm Born}=Z_{\rm LP}^2|X_{\rm LP}|^4\Gamma^{(0)}(\0,0;\0,0)=\frac{2\pi |X_{\rm LP}|^4}{m_r},
\end{align}
where the last term is evaluated using the expressions in Appendix~\ref{app:appendix}. Here, the factor $\frac{2\pi}{m_r}$ is the Born approximation for the exciton-exciton
interaction constant in the absence of light-matter coupling.
We can directly compare this result with the Born approximation in the case of Coulomb interaction, $g_{\rm PP}^{\rm Born}\simeq \frac{3.03|X_{\rm LP}|^4}{m_r}$~\cite{TassonePRB99}. 
We see that, in the Born approximation, the P-P interaction constant is roughly a factor two larger within our model~\eqref{eq:Hamiltonian} than in the case of Coulomb interactions between charge carriers.

\subsection{Calculation of the polariton-polariton $T$ matrix}\label{subsec:numerical_implementation}

As seen in Eq.~\eqref{eq:gPP}, the interaction coefficient $g_{\rm PP}$ is evaluated for polaritons at zero momentum, and hence it only has a contribution from $s$-wave scattering. We thus start by defining the $s$-wave components of $T$, $\Gamma$, and $\chi$ as
\begin{align}
    T_s(E;p_1,\epsilon_1;p_2,\epsilon_2)&=\int_0^{2\pi} \frac{d\phi_{12}}{2\pi} T(E;\p_1,\epsilon_1;\p_2,\epsilon_2),\\
    \Gamma_s(p_1,\epsilon_1;p_2,\epsilon_2)&=
    \int_0^{2\pi} \frac{d\phi_{12}}{2\pi}\Gamma(\p_1,\epsilon_1;\p_2,\epsilon_2),\\
    \chi_s(p_1,\epsilon_1;\Q_1,\Q_2)&=
    \int_0^{2\pi} \frac{d\phi_{1}}{2\pi}\chi(\p_1,\epsilon_1;\Q_1,\Q_2),
\end{align}
where we have taken advantage of the fact that the system is isotropic, i.e., the scattering does not depend on the overall orientation but only on the angle between incident and outgoing particles. Here, $\phi_i$ is the angle of $\p_i$ with respect to a fixed axis, and $\phi_{12}=\phi_1-\phi_2$ is the angle between $\p_1$ and $\p_2$. While $\chi_s$ does not depend on the overall orientation of the system, it still depends on the angle between $\Q_1$ and $\Q_2$. With these $s$-wave projections, we can solve the set of equations \eqref{eq:T-matrix_full}-\eqref{eq:chi_full} fully within the $s$-wave channel.

To proceed, we discretize all integrals using Gauss-Legendre quadrature. In this way, we convert all integral equations into sets of linear equations that can be solved by appropriate matrix inversions~\cite{Numerical_Recipes_Integral_Equations}. In defining our grids, we use a Wick rotation for the energy shift in $T$, which means that $\epsilon_i$ goes from $-i\infty$ to $+i\infty$~\cite{LevinsenPRA06}, effectively shifting the integration contour away from poles and branch cuts of the polariton propagators. We then solve for $\chi$ on this grid using Eq.~\eqref{eq:chi_full}, insert the result into the equation for $\Gamma$, Eq.~\eqref{eq:Gamma_full}, and finally obtain $T$ by solving Eq.~\eqref{eq:T-matrix_full}. In the final step, we use the fact that $T$ only depends parametrically on incoming momentum and energy shift to solve for $T_s(2E_{\rm LP};0,0;p_2,\epsilon_2)$, and we then iterate Eq.~\eqref{eq:T-matrix_full} to obtain $T_s(2E_{\rm LP};0,0;0,0)$.

Finally, we remark on what low-energy scattering of polaritons means in practice. Two-dimensional scattering theory dictates that the two-body scattering $T$ matrix must vanish when the particles' momenta vanish~\cite{AdhikariAJP86}. However, due to the light-matter coupled nature of polaritons, the polariton $T$ matrix only goes to zero below the momentum scale $k^*\sim e^{-m_X/m_c}/a_0$, with $a_0=1/\sqrt{2m_r\eb}$ the exciton Bohr radius~\cite{Bleu2020}. Given the large exciton-photon mass ratio, this momentum scale is so small that it is inaccessible in any realistic experiment. Therefore, $p_1=p_2=0$ should be understood as $k^*\ll p_1,p_2\ll 1/a_0$.

\subsection{Polariton-polariton interaction strength}\label{subsec:numerical_results}

\begin{figure}[t!]
	\centering
	\includegraphics[width=\linewidth]{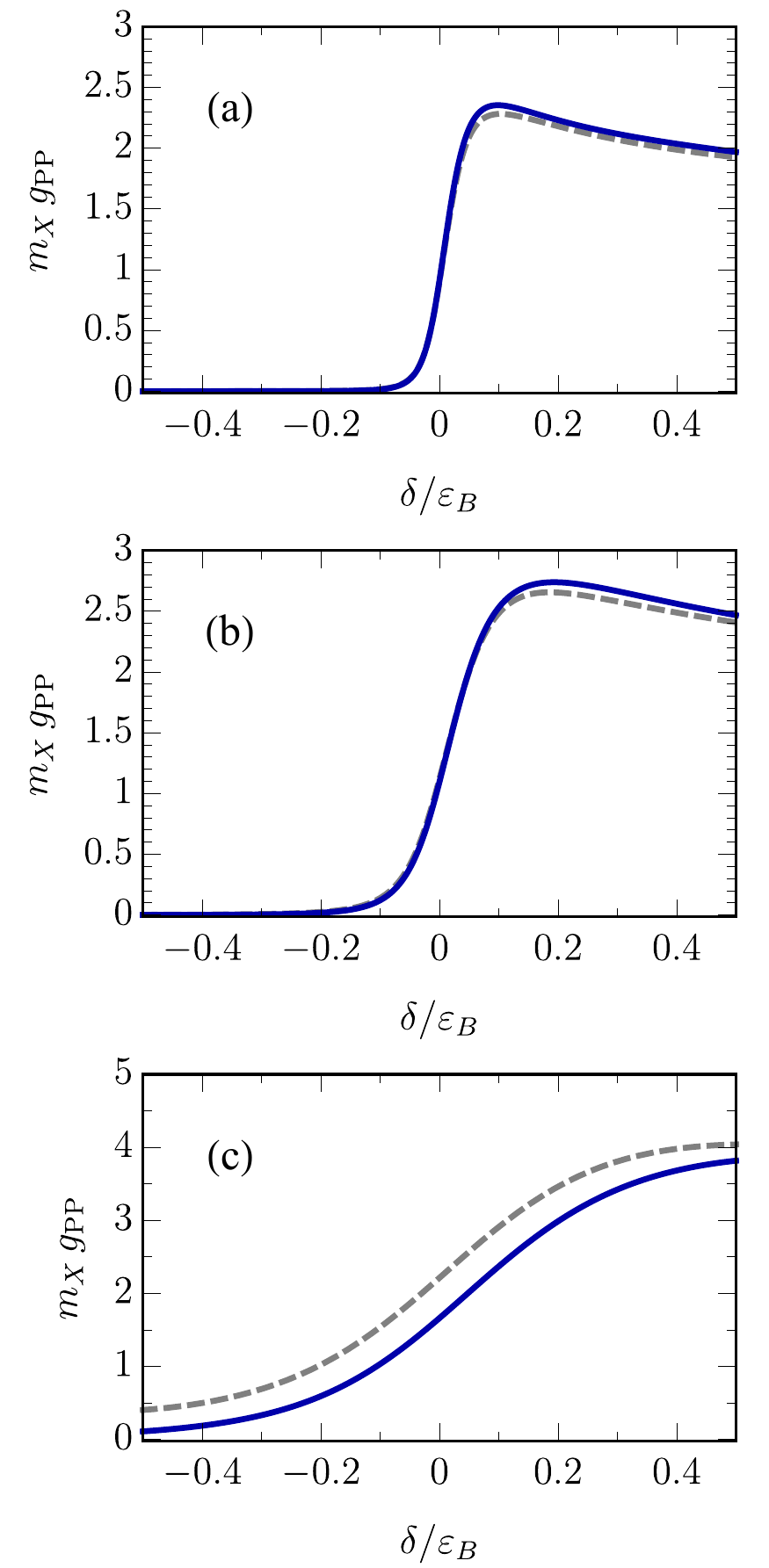}
	\caption{Polariton-polariton interaction constant $g_{\PP}$ (blue solid line) as a function of detuning. We show our results for a fixed Rabi coupling of (a) $\Omega/\varepsilon_B=0.025$ corresponding approximately to the case of a MoSe$_2$~\cite{dufferwiel2015exciton}, MoS$_2$~\cite{liu2015strong}, or WSe$_2$ \cite{lundt2016room,He2014} monolayer; (b) $\Omega/\varepsilon_B=0.05$ corresponding to a WS$_2$~\cite{Flatten2016} monolayer; (c) $\Omega/\varepsilon_B=0.2$. In all panels, we have $m_e/m_h=1$~\cite{TMDREV2018} and $m_c/m_X=10^{-4}$. The grey dashed line is the approximation in Eq.~\eqref{eq:Oli2}.}
	\label{fig:T0_plot}
\end{figure}

\subsubsection{Transition metal dichalcogenides}
In Fig.~\ref{fig:T0_plot}, we show the result of our numerically exact calculation of the P-P interaction constant as a function of detuning for $m_e = m_h$, with parameters corresponding to a monolayer TMD [panels (a,b)], and with a larger value of the light-matter coupling relative to the exciton binding energy [panel (c)]. In all cases, we find that the interaction increases as we go from photon-dominated polaritons at negative detuning toward positive detuning, which is as expected. However, we also find that the interaction constant has a peak at positive detuning, a feature which is absent in the commonly applied Born approximation, Eq.~\eqref{eq:Born}. Furthermore, comparing panels (a) to (c), we see that the interaction generally increases with increasing light-matter coupling $\Omega/\eb$.

Figure~\ref{fig:T0_plot} also compares our diagrammatic calculation with the universal form of the low-energy P-P scattering introduced in Ref.~\cite{Bleu2020}:
\begin{align} \label{eq:Oli2}
    g_{\PP}\simeq\frac{4\pi \abs{X_{\rm LP}}^4}{m_X\ln\left(\frac{\varepsilon_X}{2\abs{E_{\rm LP}+\eb}}\right)}.
\end{align}
We see that this simple approximation is highly accurate  
when $\Omega/\eb\ll1$, which is to be expected since it is based on low-energy exciton-exciton  (X-X)
scattering. We stress that Eq.~\eqref{eq:Oli2} only contains one parameter that characterizes the low-energy X-X scattering: the energy scale $\varepsilon_X$. This is in turn related to the 2D X-X 
$s$-wave scattering length $a_X$ via $\varepsilon_X\equiv\frac{1}{m_X\,a_X^2}$. Within our description of the electronic interactions, this scattering length was previously calculated in the context of ultracold atoms to be $a_X\approx0.56a_0$~\cite{Petrov2003}, which we have used to generate the corresponding lines. This value has also been obtained by Quantum Monte Carlo calculations~\cite{Bertaina2011} and from mean-field theory with Gaussian fluctuations~\cite{He2015}. 
From the form of Eq.~\eqref{eq:Oli2}, we see that there is a competition between 
the Hopfield coefficient and the logarithm, since the excitonic fraction $|X_{\rm LP}|^2$ increases towards 1 as the polariton energy approaches that of the exciton, while the logarithm diverges as $E_{\rm LP} \to -\eb$, thus suppressing the scattering process. 
This explains the presence of the peak at positive detuning in Fig.~\ref{fig:T0_plot}. 
We note that a low-energy expression similar to Eq.~\eqref{eq:Oli2} has already been verified in the case of polariton-electron scattering by comparing with exact three-body calculations \cite{Li2021PRL,Li2021PRB}. 

\begin{figure}[t!]
	\centering
	\includegraphics[width=\linewidth]{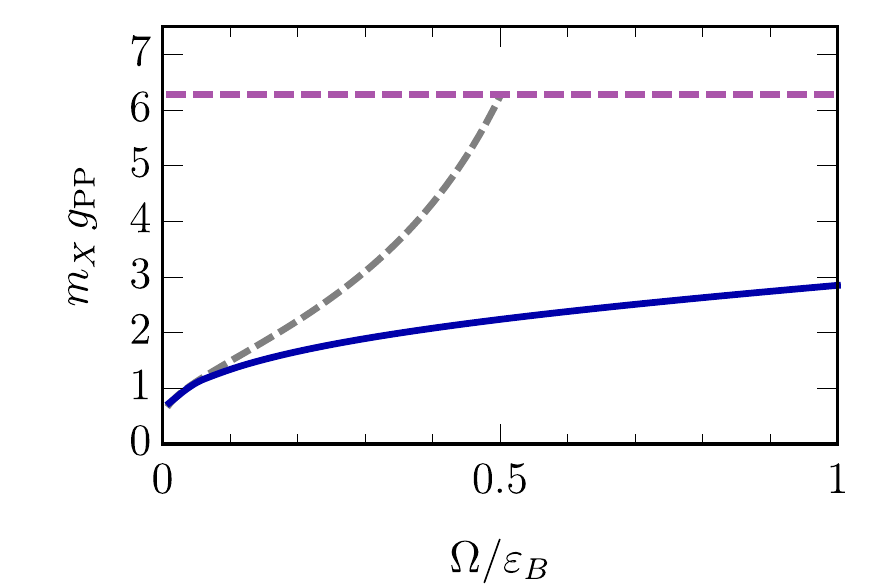}
	\caption{Polariton-polariton interaction constant $g_{\PP}$ as a function of Rabi coupling $\Omega$, at a fixed detuning of $\delta=0$. As in Fig.~\ref{fig:T0_plot}, we take $m_e/m_h=1$ and $m_c/m_X=10^{-4}$. We show the exact four-body calculation (blue solid line), the approximation in Eq.~\eqref{eq:Oli2} (grey dashed line), and the Born approximation in Eq.~\eqref{eq:gPPBorn} (purple dashed line).}
	\label{fig:Omega_sweep}
\end{figure}

Our results also show that the interaction constant is significantly smaller than that predicted by the Born approximation, which, for instance, yields an interaction constant $g_{\rm PP}^{\rm Born}=2\pi/m_X$ at zero detuning according to Eq.~\eqref{eq:gPPBorn}, independently of the strength of the light-matter coupling~\footnote{In practice, within the model~\eqref{eq:Hamiltonian} there is a small correction to the Hopfield coefficients at large Rabi coupling~\cite{Li2021PRB}, which we neglect here.}. The qualitative difference between our results and the Born approximation is particularly evident when considering the system at a fixed exciton fraction as a function of light-matter coupling. This is shown in Fig.~\ref{fig:Omega_sweep} where we plot the interaction constant as a function of Rabi coupling (which corresponds 
to scanning across different semiconductor microcavities) with fixed detuning $\delta=0$ such that the polariton is half light, half matter. We see that the interaction becomes stronger with increasing Rabi coupling, which we interpret as being due to the stronger energy shift of the polariton from the exciton. When $\Omega\ll\eb$, we see that the Born approximation greatly overestimates the interactions, which only approach $g_{\rm PP}^{\rm Born}$ when $\Omega\gtrsim\eb$ (note that the Born approximation serves as an upper bound on the interaction constant when there are no biexciton bound states~\cite{Li2021PRB}). 
On the other hand, the universal low-energy expression in Eq.~\eqref{eq:Oli2} is clearly seen to work well at small $\Omega/\eb$. 

It is important to remember that Eq.~\eqref{eq:Oli2} will no longer be valid at large Rabi coupling, $\Omega\sim\eb$, and/or at large negative detuning $\delta\sim-\eb$~\cite{Bleu2020}. Both of these cases are formally outside the regime of validity of a model assuming structureless excitons since the assumption of low-energy exciton-exciton 
scattering is no longer satisfied. In this limit, Eq.~\eqref{eq:Oli2} predicts a divergence when $2\abs{E_{\rm LP}+\eb}=\varepsilon_X$. While this divergence has previously been used to argue that the polariton interactions could be anomalously enhanced at large negative detuning~\cite{hu2020twodimensional}, we see that this divergence is cured by considering the full four-body problem, and it is therefore an unphysical feature of the approximation. We also note that such a divergence is incompatible with the fact that the Born approximation serves as an upper bound of the interaction constant --- see Fig.~\ref{fig:Omega_sweep}.

\begin{figure}
	\centering
	\includegraphics[width=0.95\linewidth]{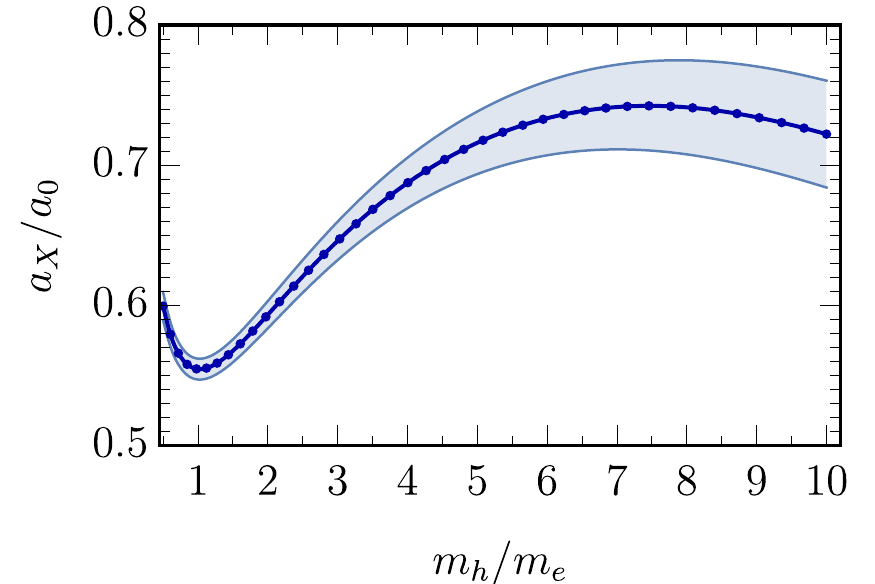}
	\caption{Exciton-exciton scattering length $a_X$ as a function of inverse electron-hole mass ratio. Blue circles are the best fits obtained from the procedure described in the main text, the blue line is a guide to the eye through the data points, and the shaded area represents the uncertainty of the fit.}
	\label{fig:me_sweep}
\end{figure}

\subsubsection{Systems with electron-hole mass imbalance}
Next we consider the effect of changing the electron-hole mass ratio $m_e/m_h$. Physically, this corresponds to different semiconductor microcavities. For instance, in a GaAs quantum well we have $m_e/m_h\approx 0.15$ \cite{GaAsEffectiveMass}. 
Figure~\ref{fig:me_sweep} shows the fitted value of the X-X scattering length $a_X$ as a function of electron-hole mass ratio. Each data point was obtained by solving the four-body $T$-matrix equation \eqref{eq:T-matrix_full} with a fixed $m_c/m_X=10^{-4}$ and $\Omega/\varepsilon_B=0.05$ within the detuning range of $-0.1\leq\delta/\varepsilon_B\leq 0.1$, and then fitting Eq.~\eqref{eq:Oli2} using $a_X=1/\sqrt{m_X\varepsilon_X}$ as a free parameter. This fitting procedure builds on the observation in  Fig.~\ref{fig:T0_plot} that the simple expression overlaps mostly with the four-body calculation in the range of small detuning. We see that the scattering length increases in mass-imbalanced systems, corresponding to increased polariton-polariton interactions.

\begin{figure}[t!]
	\centering
	\includegraphics[width=0.95\linewidth]{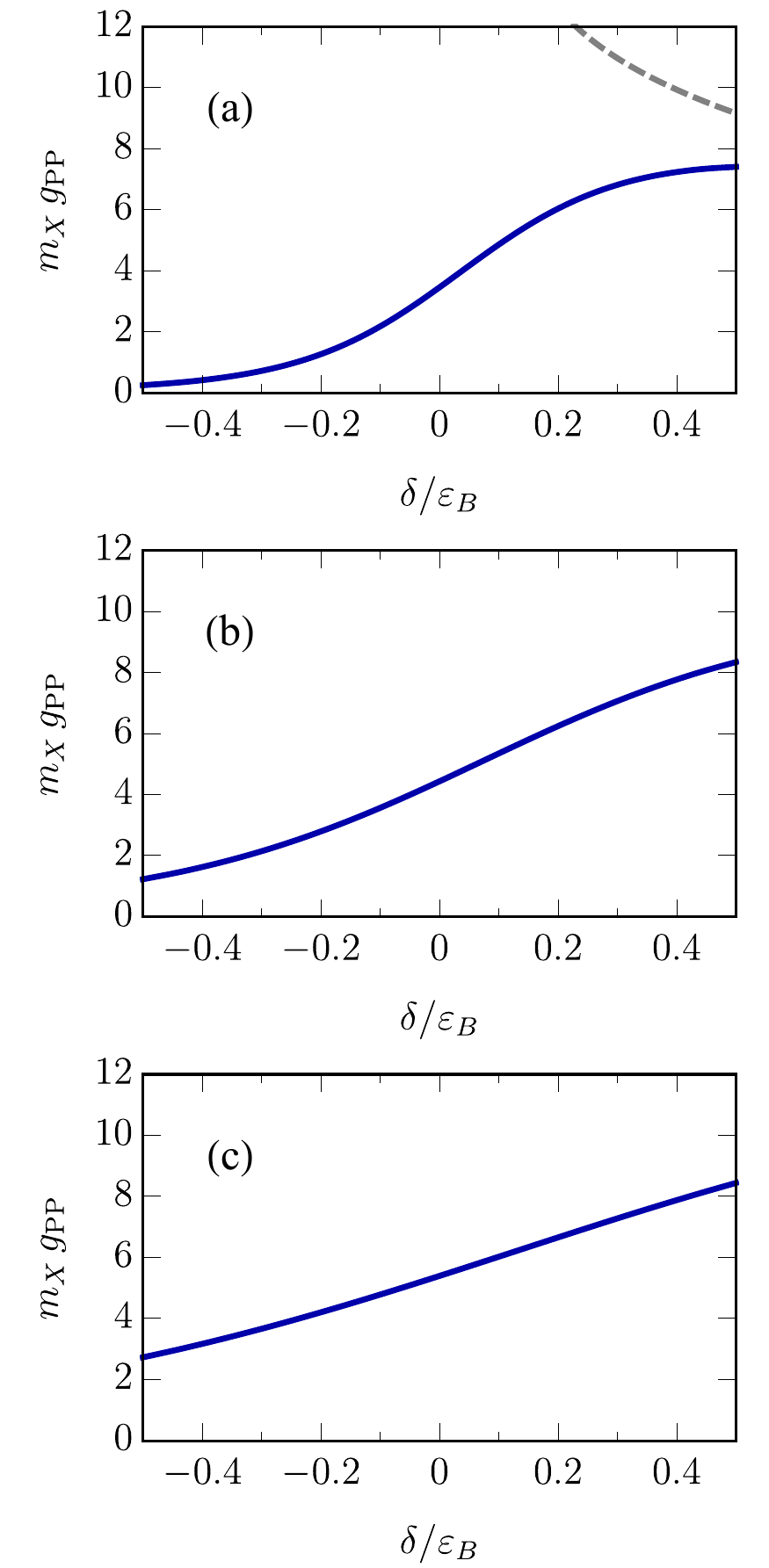}
	\caption{Polariton-polariton interaction constant $g_{\PP}$ (blue solid line) in a GaAs system with varying Rabi coupling
	 (a) $\Omega/\varepsilon_B=0.2$; (b) $\Omega/\varepsilon_B=0.2\sqrt{4}=0.4$; (c) $\Omega/\varepsilon_B=0.2\sqrt{12}\approx 0.7$. 
	 In all panels, we have $m_e/m_h= 0.15$ and $m_c/m_X=10^{-4}$. The grey dashed line is the approximation in Eq.~\eqref{eq:Oli2} with the value $a_X/a_0=0.74$ obtained from Fig.~\ref{fig:me_sweep}.}
	\label{fig:T0_plot_GaAs}
\end{figure}

For the particular case of a GaAs quantum well system, the electron-hole mass ratio $m_e/m_h\approx0.15$ generally implies that the value of $g_{\PP}$ is increased compared to the corresponding equal-mass case. We show our results in Fig.~\ref{fig:T0_plot_GaAs} for three different Rabi couplings, corresponding to systems with 1, 4, or 12 quantum wells (however, we stress that these results are still calculated within our model Hamiltonian~\eqref{eq:Hamiltonian} which considers a single quantum well; we will discuss the extension of our theory to multilayer systems below in Sec.~\ref{sec:multi}). Indeed, we see that the result in Fig.~\ref{fig:T0_plot_GaAs}(a) is larger than that for equal masses in Fig.~\ref{fig:T0_plot}(c), with all other parameters being the same. We also see that, due to the large value of $a_X$ for this mass ratio, the simple approximation in Eq.~\eqref{eq:Oli2} breaks down for smaller Rabi couplings, and even for the Rabi coupling corresponding to a single GaAs quantum well the approximation only works for large positive detuning [Fig.~\ref{fig:T0_plot_GaAs}(a)].

\begin{figure}
	\centering
	\includegraphics[width=\linewidth]{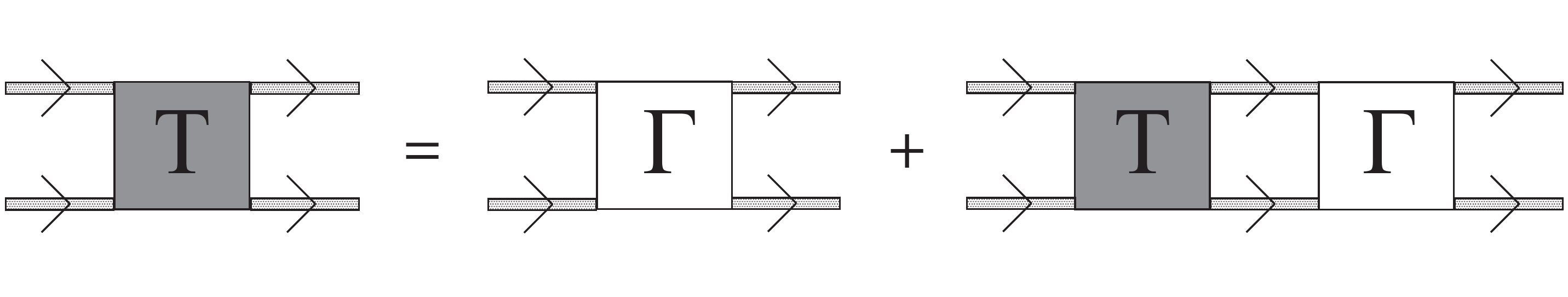}
	\caption{The polariton-polariton scattering process can be approximated as an off-shell exciton-exciton scattering process. The exciton propagators are represented by shaded rectangles with arrows. The $\Gamma$ diagram is given by Fig.~\ref{fig:Gamma_and_chi} and Fig.~\ref{fig:Gamma(0)} by replacing the polariton propagators with the exciton ones.}
	\label{fig:T-matrix_XX}
\end{figure}

\begin{figure*}[t!]
	\centering
	\includegraphics[width=\linewidth]{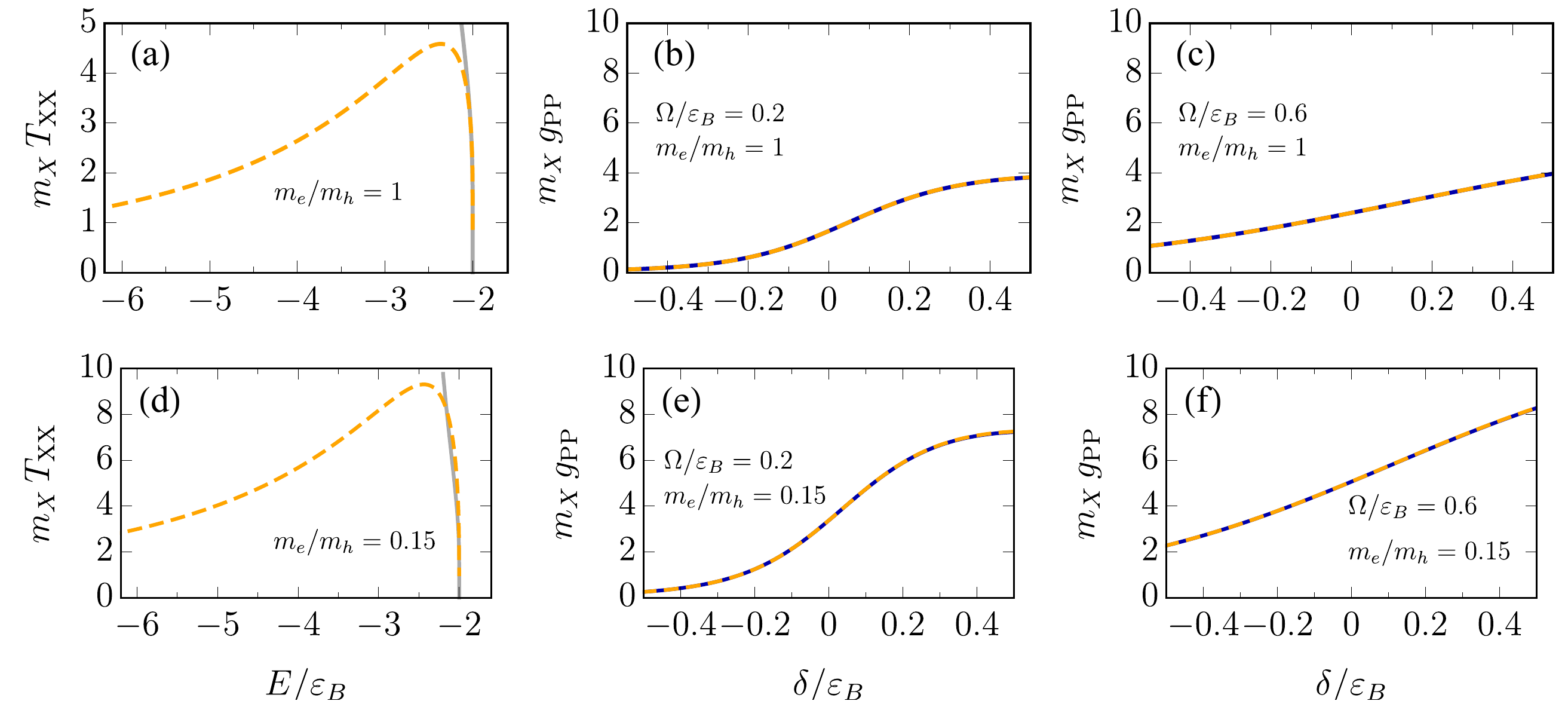}
	\caption{(a,d) Exciton-exciton $T$ matrix at negative energy (yellow dashed curve) and the low-energy approximation in Eqs.~\eqref{eq:TXX} and \eqref{eq:delta} (gray solid curve). (b,c,e,f) Comparison of the polariton-polariton interaction constant of the full four-body calculation (blue solid curve) and the off-shell exciton-exciton scattering approximation Eq.~\eqref{eq:gPPXX} (yellow dashed curve) for various Rabi couplings. (a-c) correspond to TMD mass ratio, (d-f) correspond to GaAs mass ratio.}
	\label{fig:PP_4body_XXshift_Compare}
\end{figure*}

\subsection{Off-shell exciton-exciton scattering approximation}\label{sec:XX_energy_shift}

We have seen in Fig.~\ref{fig:T0_plot} that the universal low-energy formula in Eq.~\eqref{eq:Oli2} works extremely well at relatively small $\Omega/\eb$, while Fig.~\ref{fig:T0_plot_GaAs} illustrated its breakdown at stronger light-matter coupling and in the mass-imbalanced system. We now discuss how we can generalize Eq.~\eqref{eq:Oli2} to stronger light-matter coupling. The central point is that because the underlying interactions are electronic, the typical momenta relevant to the scattering process are $k\sim 1/a_0$. At these momenta, the lower polariton propagator in Eq.~\eqref{eq:Dp} essentially reduces to the exciton propagator in Eq.~\eqref{eq:D0} due to the small photon mass. Therefore, to a large degree of accuracy, we can replace all internal polariton propagators with exciton propagators in Eqs.~\eqref{eq:T-matrix_full}-\eqref{eq:chi_full}. Formally, this corresponds to taking the limit of vanishing photon mass, $m_c\to0$.

Consequently, we can directly relate our results to the corresponding exciton-exciton scattering problem, which is outlined in Appendix~\ref{app:appendixXX} and illustrated in Fig.~\ref{fig:T-matrix_XX}. Taking into account the difference in the overall normalization of the $T$ matrices in the X-X and the P-P 
scattering [compare Eqs.~\eqref{eq:Tfromt} and \eqref{eq:T-matrix_fullXX}], we find
\begin{align}\label{eq:gPPXX}
    g_{\rm PP}\simeq |X_{\rm LP}|^4\left(\frac{E_{\rm LP}}{\eb}\right)^2   T_{\rm XX}(2E_{\rm LP}).
\end{align}
Here, $T_{\rm XX}(2E_{\rm LP})$ is the X-X scattering $T$ matrix evaluated for excitons at rest with a shifted collision energy. 
Equation~\eqref{eq:gPPXX} has a natural interpretation as an off-shell exciton scattering process weighted by the normalization of the external polariton propagators. The off-shell exciton approximation provides a different perspective in understanding the polariton-polariton interaction process. As we show in the next sections, this will be useful in understanding the polariton-exciton interaction process, and in generalizing our results to multilayer systems. We furthermore expect that this result can be generalized to scattering processes involving other light-matter coupled quasiparticles.

We illustrate the utility of Eq.~\eqref{eq:gPPXX} in Fig.~\ref{fig:PP_4body_XXshift_Compare}. Panels (a) and (d) show the X-X scattering $T$ matrix as a function of energy for the case of $m_e=m_h$ and $m_e=0.15m_h$ relevant to TMDs and GaAs quantum wells, respectively. This function then serves as an input that allows us to approximate, for instance, all the results in Fig.~\ref{fig:T0_plot}. In the remaining panels we show the comparison between our numerically exact results and the approximation in Eq.~\eqref{eq:gPPXX} for large Rabi couplings, where the simple expression in Eq.~\eqref{eq:Oli2} clearly fails. We see that the agreement is essentially perfect, with the relative error incurred of order $m_c/m_X$. This clearly demonstrates that 
P-P scattering should be understood as off-shell X-X scattering.

Finally, we comment on the relationship between the two approximations in  Eqs.~\eqref{eq:Oli2} and~\eqref{eq:gPPXX}. The X-X $T$ matrix at energy $E$ takes the universal form~\cite{LevinsenBook15}
\begin{align}\label{eq:TXX}
    T_{\rm XX}(E)= \frac{-4/m_X}{\cot\delta(E_{\rm coll})-i},
\end{align}
in terms of the scattering phase shift $\delta$, where the collision energy $E_{\rm coll} = E+2\eb$, since the energy $E$ is measured from the electron-hole continuum. The phase shift, in turn, has the universal low-energy expansion~\cite{AdhikariAJP86}
\begin{align}\label{eq:delta}
    \cot\delta(E_{\rm coll})\simeq \frac1\pi\ln\left[\frac{E_{\rm coll}+i0}{\varepsilon_X}\right]+O(E_{\rm coll}),
\end{align}
which is valid when $|E_{\rm coll}| \ll \eb$, i.e., when  the energy is close to that of two excitons. 
Indeed, panels (a) and (d) of Fig.~\ref{fig:PP_4body_XXshift_Compare} show that, in this regime, Eqs.~\eqref{eq:TXX} and \eqref{eq:delta} agree very well with our exact calculation of $T_{\rm XX}$.
Keeping the leading term of the phase shift and taking $E_{\rm LP}\lesssim-\eb$ in the prefactor, Eq.~\eqref{eq:gPPXX} exactly reduces to Eq.~\eqref{eq:Oli2}.

\section{Polariton-exciton Interaction}\label{sec:PX}

\begin{figure}[h]
	\centering
	\includegraphics[width=\linewidth]{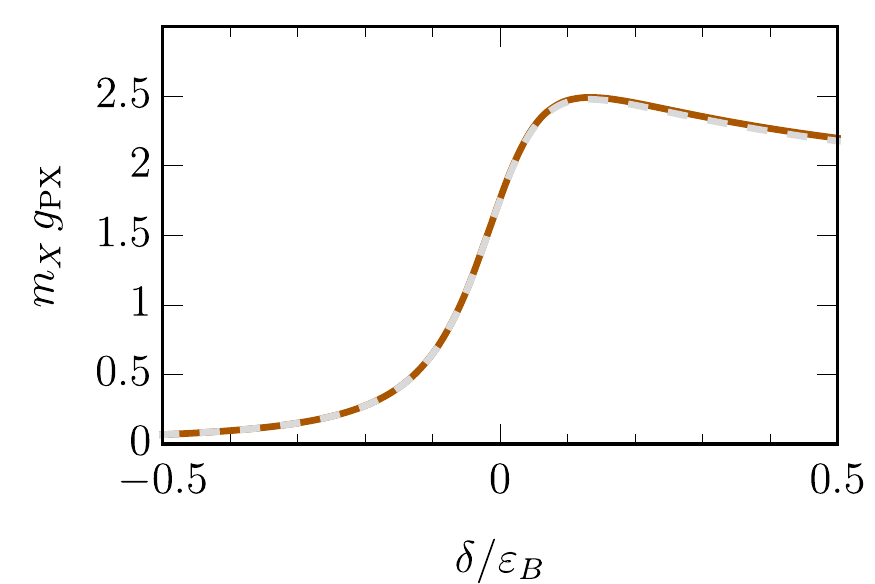}
	\caption{Comparison of four-body calculation Eq.~\eqref{eq:gPX_4-body} (brown solid curve) and the approximation in Eq.~\eqref{eq:gPXXX} (grey dashed curve) for the polariton-exciton interaction constant $g_{\rm PX}$ with $\Omega/\varepsilon_B=0.05$, $m_e/m_h=1$, and $m_c/m_X=10^{-4}$ corresponding to a WS$_2$ monolayer.}
	\label{fig:PX_T0}
\end{figure}

The polariton-exciton (P-X) interaction constant $g_{\PX}$ can be calculated using the same diagrammatic method as that applied in Sec.~\ref{sec:PP} to polaritons. In this case, we consider the incident/outgoing particles to be one exciton and one polariton. Since we are working with a single semiconductor layer and since we ignore any spin degree of freedom, there is only one type of exciton. Hence, the exciton state couples to light, and all internal electron-hole pair propagators will be polaritonic. Therefore, with the exception of external legs, the diagrams are unchanged from those in Figs.~\ref{fig:PP_T-matrix} and \ref{fig:Gamma_and_chi}. Note that this situation can be more complicated in microcavities with multiple semiconductor layers/quantum wells; we discuss this scenario below in Sec.~\ref{sec:multi}.

To be concrete, we will consider the low-energy scattering of a polariton with an exciton in the limit where both of these are at rest. The particles have rest energies $E_{\rm LP}$ and $-\eb$, respectively, and thus the total energy is $E=E_{\rm LP}-\eb$. To ensure the correct rest energies in our calculation, we therefore take $\epsilon_1=\epsilon_2=(E_{\rm LP}+\eb)/2$ at the end of the calculation~\footnote{Note that we can still perform a Wick rotation in $\epsilon_i$ in our integral equation; we simply perform the rotation in the complex plane around the point $(E_{\rm LP}+\eb)/2$.}. With these considerations, we arrive at the equation for the P-X scattering $T$ matrix:
\begin{align}
	T_{\rm PX}(E;\p_1,\epsilon_1;\p_2, 
	\epsilon_2)=&Z_X\sqrt{Z_-(\p_1)Z_-(\p_2)}\nn \\ &\hspace{-20mm}
	\times |X_-(\p_1)X_-(\p_2)|t(E;\p_1,\epsilon_1;\p_2, \epsilon_2),
\end{align}
in terms of the unnormalized $T$ matrix in Eq.~\eqref{eq:T-matrix_full}.
Here, the prefactor 
is due to the normalization of external legs, and we have defined the exciton residue $Z_X\equiv2\pi\eb/m_r$~\cite{Li2021PRB}. The $T$ matrix allows us to obtain the P-X interaction constant
\begin{align}
    g_{\rm PX}=T_{\rm PX}(E_{\rm LP}-\eb;\0,(E_{\rm LP}+\eb)/2;\0,(E_{\rm LP}+\eb)/2).\label{eq:gPX_4-body}
\end{align}
This interaction constant would appear, for instance, in a Gross-Pitaevskii treatment of a polariton condensate interacting with an excitonic reservoir~\cite{WouterPRL07}. We also expect this to be relatively insensitive to the momentum $\p$ of the scattering exciton if $|\p| \ll \sqrt{m_X \Omega}$ such that the collision energy is unaffected.

Figure~\ref{fig:PX_T0} displays the result of our calculation for the case of a TMD monolayer. We see that qualitatively the result is similar to the P-P interaction constant, with a maximum close to $\delta=0$. Furthermore, we note that contrary to the expectations of the Born approximation we can have $g_{\rm PX}\lesssim g_{\rm PP}$ when $\delta$ is above this maximum, since in this regime the strength of interactions is dominated by the energy shift due to the light-matter coupling, which is larger in the two-polariton problem.

By the same reasoning as in Sec.~\ref{sec:XX_energy_shift}, we can approximate all internal polariton propagators in the collision process by their excitonic counterparts. Since the excitonic system is Galilean invariant, it only depends on the total energy in the center-of-mass frame, and therefore we can approximate
\begin{align}\label{eq:gPXXX}
    g_{\rm PX}\simeq|X_{\rm LP}|^2 \frac{\abs{E_{\rm LP}}}{\eb}   T_{\rm XX}(E_{\rm LP}-\eb).
\end{align}
Figure~\ref{fig:PX_T0} shows the essentially perfect agreement between this result and the exact calculation in Eq.~\eqref{eq:gPX_4-body}. This is highly significant, since it implies that in general we can extract $g_{\rm PX}$ directly from our calculation of $T_{\rm XX}$ in Fig.~\ref{fig:PP_4body_XXshift_Compare}(a), and likewise for the GaAs mass ratio in Fig.~\ref{fig:PP_4body_XXshift_Compare}(d).

Finally, by using the low-energy expression in Eqs.~\eqref{eq:TXX} and \eqref{eq:delta}, we find the universal low-energy behavior
\begin{align}\label{eq:PX_Olivier_expression}
    g_{\PX}\simeq\frac{4\pi \abs{X_{\rm LP}}^2}{m_X\ln\left(\frac{\varepsilon_X}{\abs{E_{\rm LP}+\eb}}\right)},
\end{align}
valid when $|E_{\rm LP}+\eb|\ll \eb$. This proves the corresponding expression in Eq.~\eqref{eq:Olivier_expression}.

\section{Implications for multilayer microcavities}
\label{sec:multi}

\begin{figure}
	\centering
	\includegraphics[width=\linewidth]{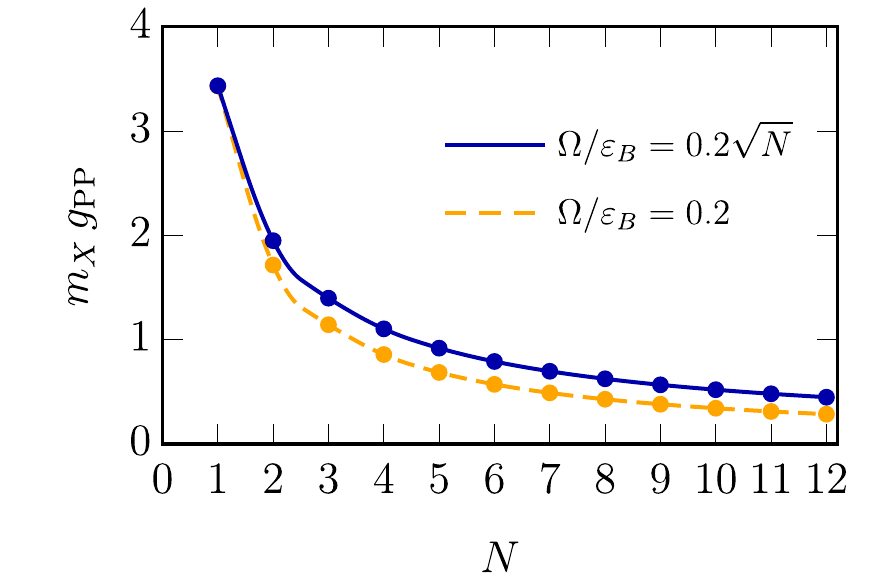}
	\caption{Comparison of the polariton-polariton interaction constant in a $N$-layer GaAs system with $m_e/m_h=0.15$ and $m_c/m_X=10^{-4}$, calculated by different methods. Solid dots are the calculated data points and lines are guides to the eye.
	The blue-solid curve is our expression for the multilayer interaction, Eq.~\eqref{eq:gPPXXNapprox}, at $\delta=0, \,\Omega/\eb=0.2\sqrt{N}$. The yellow-dashed curve is the single layer four-body result divided by $1/N$ at $\delta=0, \,\Omega/\eb=0.2$.}
	\label{fig:GaAs_N_layer}
\end{figure}

Thus far, we have been focussing on systems with a single semiconductor layer in a microcavity. This could, for instance, be a TMD monolayer or a single GaAs quantum well. However, in many situations it is advantageous to use structures with multiple semiconductor layers, since the Rabi coupling increases with the number of layers. Without writing down an explicit Hamiltonian for the multilayer system, we will now take advantage of our observation that polaritons interact as off-shell excitons to discuss how our results generalize to the multilayer case.

We therefore consider polariton scattering in a structure with $N$ layers which are equally coupled to light, but which are otherwise independent in the sense that there is no tunnelling between them. In this case, the coupling to light defines a bright exciton, an equal superposition of excitons in each semiconductor layer, and $N-1$ dark excitons that remain uncoupled~\cite{CiutiRevMod13}. The effective Rabi coupling of the bright exciton is now enhanced to $\Omega=\sqrt{N}\Omega_1$, where $\Omega_1$ is the Rabi coupling to a single layer. Within the new single-particle basis of bright and dark excitons, the interaction term in the Hamiltonian is rather complicated~\cite{Bleu2020}. However, the key idea is that the underlying electronic interactions in polariton scattering only take place within the individual layers. Since each exciton in a particular layer within a polariton has an amplitude $1/\sqrt{N}$, and since there are $N$ semiconductor layers, the polariton-polariton scattering $T$ matrix is suppressed by an overall factor $1/N$. 
Therefore, taking into account the energy shift in the scattering process due to the light-matter coupling and using our previous observation that the polariton interaction can very accurately be viewed as off-shell exciton scattering, we find the simple expression for the multilayer P-P interaction constant
\begin{align}
    g_{\rm PP}(N)&
    \simeq \frac1N|X_{\rm LP}|^4\left(\frac{E_{\rm LP}}{\eb}\right)^2   T_{\rm XX}(2E_{\rm LP}).
    \label{eq:gPPXXNapprox}
\end{align}
Importantly, apart from  the factor $1/N$, the right hand side corresponds precisely to the polariton interaction constant for a single semiconductor layer in Eq.~\eqref{eq:gPPXX}, evaluated with the Rabi coupling \textit{for the $N$-layer system}. 
While this equation may look like a trivial extension to $N$ layers, we remind the reader that the interaction constant in a single layer generically increases with Rabi coupling, as shown in Fig.~\ref{fig:Omega_sweep}.
To illustrate this effect, in Fig.~\ref{fig:GaAs_N_layer} we show our results for $g_{\rm PP}(N)$ as a function of $N$ in the case of a GaAs quantum well microcavity, where up to 12 quantum wells are routinely used. While we indeed see that the suppression with $N$ is dominant, the result of Eq.~\eqref{eq:gPPXXNapprox} is 50$\%$ higher for 12 quantum wells than what would be expected based on a simple $1/N$ scaling.

\begin{figure}
	\centering
	\includegraphics[width=\linewidth]{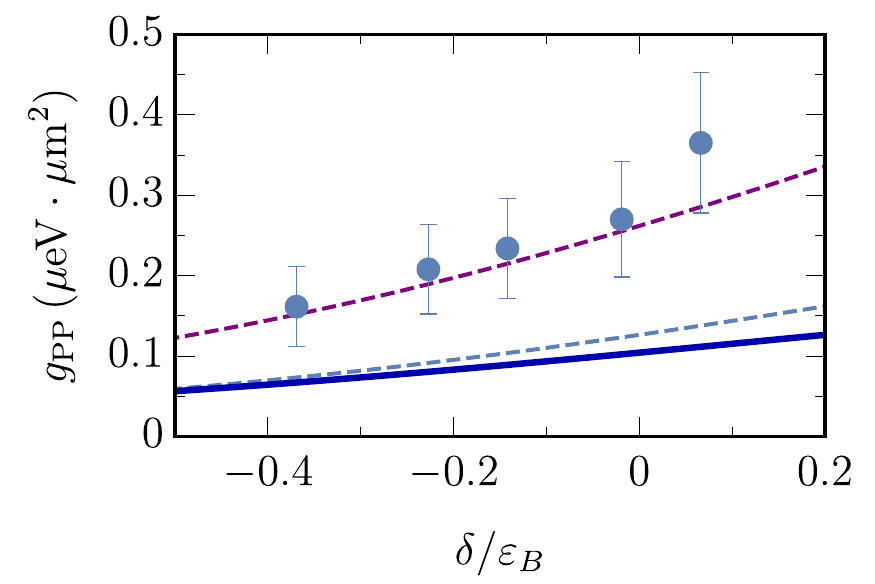}
	\caption{Comparison between experimental measurement of the polariton-polariton interaction constant extracted from the polariton blueshift~\cite{EliPRB2019} and theoretical calculations (note that the experimental result is larger in Fig.~\ref{fig:Eli_data_compare} than in Ref.~\cite{EliPRB2019} by a factor two, since we do not divide by the two polariton polarizations).
	The solid curve is our four-body calculation with $\Omega/\eb=0.75$, $m_e/m_h=0.15$, and $m_c/m_X=10^{-4}$. The exciton binding energy is $\eb=10\, \mathrm{meV}$ and the Bohr radius is $a_0=10\, \mathrm{nm}$ \cite{EliPRB2019}.
	The purple dashed curve is the Born approximation from the exciton approximation: $g_{\rm PP}^{\rm Born}(N)=\frac{4\pi\abs{X_{\rm LP}}^4}{N}\,\mathrm{\mu eV \mu m^2}$. The light-blue dashed curve is the Born approximation from the exciton approximation with Coulomb interaction: $g_{\rm PP}^{\rm Born}(N)=\frac{6.06\abs{X_{\rm LP}}^4}{N}\,\mathrm{\mu eV \mu m^2}$.
	All theory curves have taken into account the total number of quantum well layers ($N=12$).
	}
	\label{fig:Eli_data_compare}
\end{figure}

We can also use Eq.~\eqref{eq:gPPXXNapprox} to directly compare with experiment. Figure~\ref{fig:Eli_data_compare} shows our full numerical four-body result along with the experimental data from Ref.~\cite{EliPRB2019}, as well as the Born approximation within our model~\eqref{eq:Hamiltonian} and within a model that employs Coulomb electronic interactions~\cite{TassonePRB99}. 
We see that all these theory lines are not far from the experimental result. While the close agreement between experiment and the Born approximation within our model is highly suggestive, we note that this approximation corresponds to an upper bound of the theory calculation. Our exact calculation is instead about a factor of two  smaller and lies quite close to the Coulomb Born approximation~\cite{TassonePRB99}. 
Reference~\cite{EliPRB2019} has argued that the  
difference between experiment and the Coulomb Born approximation could be due to 
the quasi-two-dimensional geometry of the GaAs quantum well microcavity.

The idea of off-shell scattering can also be extended to the  P-X scattering problem. Here, the exciton involved can, for instance, be one of the $N-1$ dark excitons (corresponding to an out-of-phase superposition of excitons in different layers~\cite{Bleu2020}) or it can be an exciton in one particular layer. In both cases, by the same reasoning as above, we find
\begin{align}\label{eq:gPPXX2}
    g_{\rm PX}(N)\simeq \frac1N|X_{\rm LP}|^2 \frac{\abs{E_{\rm LP}}}{\eb}  T_{\rm XX}(E_{\rm LP}-\eb).
\end{align}
Again, apart from the factor $1/N$, the result on the right hand side corresponds precisely to the single-layer result in Eq.~\eqref{eq:gPXXX} evaluated at the polariton energy determined by the enhanced multilayer Rabi coupling. 

Finally, in the limit of weak Rabi coupling relative to the exciton binding energy, we can further simplify Eqs.~\eqref{eq:gPPXXNapprox} and \eqref{eq:gPPXX2}. Using Eqs.~\eqref{eq:TXX} and \eqref{eq:delta}, we arrive at the multilayer version of Eq.~\eqref{eq:Olivier_expression},
\begin{align}\label{eq:gPPXXNapprox2}
    g_{\PP}(N)\simeq\frac{4\pi \abs{X_{\rm LP}}^4}{m_XN\ln\left(\frac{\varepsilon_X}{2\abs{E_{\rm LP}+\eb}}\right)},
\end{align}
which was first derived in Ref.~\cite{Bleu2020}. For the P-X 
interaction constant, we likewise find
\begin{align}\label{eq:gPPXXNapprox2}
    g_{\PX}(N)\simeq\frac{4\pi \abs{X_{\rm LP}}^2}{m_XN\ln\left(\frac{\varepsilon_X}{\abs{E_{\rm LP}+\eb}}\right)}.
\end{align}

\section{Conclusions and outlook}\label{sec:conclusion}

To conclude, we have performed the first exact microscopic calculation of polariton-polariton and polariton-exciton scattering involving the internal electronic structure of identical excitons.   
Our results exploit diagrammatic techniques developed in the context of ultracold atomic gases~\cite{Brodsky2005,BrodskyPRA06,LevinsenPRA06,Levinsen2011}, appropriately extended to include the light-matter coupling~\cite{Li2021PRL,Li2021PRB}. 
We expect that this diagrammatic approach will lay the foundations for more challenging four-body calculations of the polariton interaction constants within a more realistic model featuring unscreened Coulomb interactions.

In the case of weak light-matter coupling relative to the exciton binding energy, our results showed a remarkable agreement with a recently developed universal expression based on low-energy exciton-exciton interactions~\cite{Bleu2020}. This is of practical relevance, since the regime of validity of this result includes the technologically important TMD monolayers. We furthermore extended the ideas of Ref.~\cite{Bleu2020} to show that, due to the small photon-exciton mass ratio, our results could accurately be reproduced by calculating off-shell exciton-exciton scattering processes. Here the coupling to light controls the collision energy and the overall normalization, but does not explicitly modify the diagrams themselves. We used this idea to generalize our theory to systems with multiple semiconductor layers.  
This allowed us to show that our results are consistent with recent measurements of the polariton-polariton interaction constant in a GaAs quantum well microcavity featuring 12 layers~\cite{EliPRB2019}.

The observation that collision processes involving light-matter coupled quasiparticles can be thought of as off-shell matter-only scattering is highly significant. For instance, it simplifies the calculation of the corresponding interactions using more realistic electronic interactions, since the coupling to light only affects the normalization and collision energy.  
Moreover, scattering processes at negative energy and zero momentum are similar in complexity to bound state problems which are substantially simpler than finite momentum scattering~\cite{TaylorBook}. The off-shell perspective may also prove useful in translating our results to other non-linear phenomena in exciton-polariton systems such as parametric scattering processes that populate a polariton condensate~\cite{Ciuti2003}, polariton Feshbach resonances~\cite{WoutersPRB07}, and interactions with other elementary excitations in semiconductors such as phonons~\cite{Ivanov2001,Vishnevsky2011}. More broadly, we expect the off-shell description to apply to any scattering process involving quasiparticles that are part light, part matter.

\acknowledgments 
We gratefully acknowledge insightful discussions with Olivier Bleu, Eli Estrecho, Maciej Pieczarka, and Elena Ostrovskaya, as well as helpful feedback on the manuscript from Olivier Bleu. We acknowledge support from the Australian Research Council Centre of Excellence in Future Low-Energy Electronics Technologies (CE170100039). JL and MMP are also supported through the Australian Research Council Future Fellowships FT160100244 and FT200100619, respectively.


\appendix

\section{Evaluation of the fermion exchange diagram $\Gamma^{(0)}$}\label{app:appendix}

Performing the integration over the energy $\epsilon_Q$ in Eq.~\eqref{eq:Gamma(0)_ori} by Cauchy's residue theorem, we find
\begin{align}\label{eq:Gamma(0)ABC}
\Gamma^{(0)}(\p_1,\epsilon_1; \p_2,\epsilon_2)=-2\sum_\Q\frac{A }{(A^2-B^2)(A^2-C^2)}.
\end{align}
Here we define
\begin{subequations}
\begin{align}
	&A=\epsilon-\frac{Q^2}{2m_r}-\frac{p^2_2}{8m_r}-\frac{p_1^2}{8m_r}-\frac{\p_2\cdot\p_1}{4m_e}+\frac{\p_2\cdot\p_1}{4m_h},\\
	&B=\epsilon_2-\frac{\Q\cdot\p_2}{2m_e}+\frac{\Q\cdot\p_2}{2m_h}-\frac{\Q\cdot\p_1}{2m_r},\\[1em]
	&C=\epsilon_1-\frac{\Q\cdot\p_1}{2m_e}+\frac{\Q\cdot\p_1}{2m_h}-\frac{\Q\cdot\p_2}{2m_r}.
\end{align}
\end{subequations}
The angular dependence in Eq.~\eqref{eq:Gamma(0)ABC} can be simplified in the case of equal electron and hole masses, $m_e=m_h=m$, giving:
\begin{subequations}
\begin{align}
	&A=\epsilon-\frac{Q^2}{m}-\frac{p^2_2}{4m}-\frac{p_1^2}{4m},\\[1em]
	&B=\epsilon_2-\frac{\Q\cdot\p_1}{m}, \\[1em]
	&C=\epsilon_1-\frac{\Q\cdot\p_2}{m}. 
\end{align}
\end{subequations}

\section{Exciton-exciton scattering}\label{app:appendixXX}

In the absence of light-matter coupling, we obtain the exciton-exciton scattering $T$ matrix by solving the integral equation~\cite{Levinsen2011}
\begin{align}
	T_{\rm XX}(E;\p_1,\epsilon_1;\p_2, 
	\epsilon_2)=&Z_X^2\,\Gamma(\p_1,\epsilon_1;\p_2, \epsilon_2)\nonumber\\
													&\hspace{-33mm}+
													i\int \frac{d\epsilon_q}{2\pi}\sum_\q T_{\rm XX}(E; \p_1,\epsilon_1;\q, \epsilon_q)\nn \\ &\hspace{-29mm}\times D_0 (\q,\epsilon+\epsilon_q)D_0(\q,\epsilon-\epsilon_q)\Gamma(\q,\epsilon_q;\p_2,\epsilon_2),\label{eq:T-matrix_fullXX}
\end{align}
at $E=-2\eb$. Here, $Z_X\equiv2\pi\eb/m_r$ is the residue of the exciton propagator at its energy pole which is needed for normalization~\cite{Li2021PRB}. In calculating $\Gamma$ and $\chi$ via Eqs.~\eqref{eq:Gamma_full} and \eqref{eq:chi_full} we furthermore replace all polariton propagators by their excitonic counterpart.

We then obtain the $T$ matrix discussed in Sec.~\ref{sec:XX_energy_shift} by solving Eq.~\ref{eq:T-matrix_fullXX} at $E=2E_{\rm LP}$ and taking $\p_i=0$ and $\epsilon_i=0$:
\begin{align}
    T_{\rm XX}(2E_{\rm LP})\equiv T_{\rm XX}(2E_{\rm LP};\0,0;\0,0).
\end{align}
The result of this calculation is shown in Fig.~\ref{fig:PP_4body_XXshift_Compare}.

\bibliography{P_P_scattering_refs}

\end{document}